\documentclass[amsmath,floatfix,twocolumn,superscriptaddress]{revtex4-1}
\usepackage{amssymb,amsmath}
\usepackage{graphicx,array}
\usepackage{dcolumn}
\usepackage{psfrag}
\usepackage{bm}
\usepackage{color}

\begin{document}
\title{First-principles prediction of half-Heusler half-metals above room temperature}

\author{Muhammad Atif Sattar}
\affiliation{Department of Physics, The Islamia University of Bahawalpur, 63100, Bahawalpur, Pakistan} 
\affiliation{School of Materials Science and Engineering, UNSW Sydney, NSW 2052, Australia}

\author{S. Aftab Ahmad}
\affiliation{Department of Physics, The Islamia University of Bahawalpur, 63100, Bahawalpur, Pakistan}

\author{Fayyaz Hussain}
\affiliation{Department of Physics, Bahauddin Zakariya University, 60800, Multan, Pakistan}

\author{Claudio Cazorla}
\thanks{Corresponding Author}
\affiliation{School of Materials Science and Engineering, UNSW Sydney, NSW 2052, Australia}

\begin{abstract}
Half-metallicity (HM) offers great potential for engineering spintronic applications, yet only few magnetic 
materials present metallicity in just one spin channel. In addition, most HM systems become magnetically disordered
at temperatures well below ambient conditions, which further hinders the development of spin-based electronic devices.
Here, we use first-principles methods based on density functional theory (DFT) to investigate the electronic, magnetic, 
structural, mixing, and vibrational properties of $90$ $XYZ$ half-Heusler (HH) alloys ($X =$~Li, Na, K, Rb, Cs; $Y =$~V, 
Nb, Ta; $Z =$~Si, Ge, Sn, S, Se, Te). We disclose a total of $28$ new HH compounds that are ferromagnetic, vibrationally 
stable, and HM, with semiconductor band gaps in the range of $1$--$4$~eV and HM band gaps of $0.2$--$0.8$~eV. 
By performing Monte Carlo simulations of a spin Heisenberg model fitted to DFT energies, we estimate the Curie 
temperature, $T_{\rm C}$, of each HM compound. We find that $17$ HH HM remain magnetically ordered at and above 
room temperature, namely, $300 \le T_{\rm C} \le 450$~K, with total magnetic moments of $2$ and $4$~$\mu_{\rm B}$. A 
further materials sieve based on zero-temperature mixing energies let us to conclude $5$ overall promising ferromagnetic 
HH HM at and above room temperature: NaVSi, RbVTe, CsVS, CsVSe, and RbNbTe. We also predict $2$ ferromagnetic materials 
that are semiconductor and magnetically ordered at ambient conditions: LiVSi and LiVGe. 
\end{abstract}
\maketitle

\section{Introduction}
\label{sec:intro}
Half-metals (HM) with full spin polarization at the Fermi level are of great potential for spintronic applications 
\cite{felser07,zutic03}. In particular, HM are considered to be ideal for injecting spin-polarized currents into 
semiconductors \cite{bhat16,dash09} and for manufacturing electrodes for magnetic tunnel junctions and giant magnetoresistance 
devices \cite{tanaka99,hordequin98}. Half-Heusler (HH) alloys comprise a relatively large family of multifunctional 
materials with chemical formula $XYZ$ and cubic symmetry (space group $F\overline{4}3m$); the archetypal HH compound 
NiMnSb was the first half-metal material to be ever reported \cite{groot83}. The structural compatibility of HH HM with 
typical cubic semiconductors, the potential huge number of current HH HM compounds \cite{casper12}, and the possibility 
of combining different HH to form HM layered structures \cite{azadani16}, open great prospects in the field of spin-based 
electronics.   

In recent years, HH alloys have been studied extensively with computational first-principles methods \cite{martin12,cazorla17}. 
The structural simplicity, rich variety, and predictable electronic behaviour (e.g., modified Slater-Pauling rule \cite{damewood15}) 
of HH alloys convert these materials into a perfect target for automated computational searches of HM \cite{ma17,legrain17,sattar18}. 
The structural stability and electronic and magnetic properties of HH calculated at zero temperature are the usual materials 
descriptors employed to guide the theoretical searches of candidate HM. However, analysis of the corresponding magnetic properties 
at room temperature, which are crucial for the engineering of practical applications \cite{tu16,cao17}, usually are neglected due 
to the large computational load associated with first-principles simulation of thermal effects \cite{curtarolo13}. Furthermore, 
ferromagnetic (FM) spin ordering is widely assumed in such computational investigations \cite{galanakis06,wei12} regardless of 
the fact that anti-ferromagnetic (AFM) spin ordering is also possible in HM materials \cite{damewood15,leuken94,luo08,hu11}. In 
order to provide improved guidance to future experiments, it is convenient then to examine the magnetic properties of HH HM at 
$T \neq 0$~K conditions, along with their vibrational and mixing stabilities.

In this article, we investigate the electronic, magnetic, structural, mixing, and vibrational properties of $90$ $XYZ$ 
HH alloys ($X =$~Li, Na, K, Rb, Cs; $Y =$~V, Nb, Ta; $Z =$~Si, Ge, Sn, S, Se, Te) with first-principles methods 
based on density functional theory (DFT). Our selection of materials has been motivated by recent encouraging results 
reported by other authors for some similar systems \cite{damewood15,hussain18,wang17}. We introduce a simple HH spin Heisenberg 
model fitted to FM and AFM DFT energies that allows for fast and systematic monitoring of the magnetization as a 
function of temperature. We predict that a total of $17$ HH HM are vibrationally stable and remain magnetically 
ordered at and above room temperature, with total magnetic moments of $2$ and $4$~$\mu_{\rm B}$ and semiconductor~(half-metal) 
band gaps in the range of $1$--$4$~($0.2$--$0.8$)~eV. A further materials sieve based on zero-temperature mixing energies 
allows us to identify $5$ HH HM that are most promising for electronic applications. Meanwhile, a total of $21$ HH 
alloys are found to exhibit an anti-ferromagnetic ground state but all of them turn out to be metallic. We also predict $2$ 
new semiconductor FM materials that possess high thermodynamic stability and Curie temperatures. General structural, 
thermodynamic, and functional trends are identified across the $X$, $Y$, and $Z$ series. Hence, our computational study 
discloses a number of electronic materials and design strategies that should be useful for spintronic applications. 

\begin{figure*}
\centerline{
\includegraphics[width=1.0\linewidth]{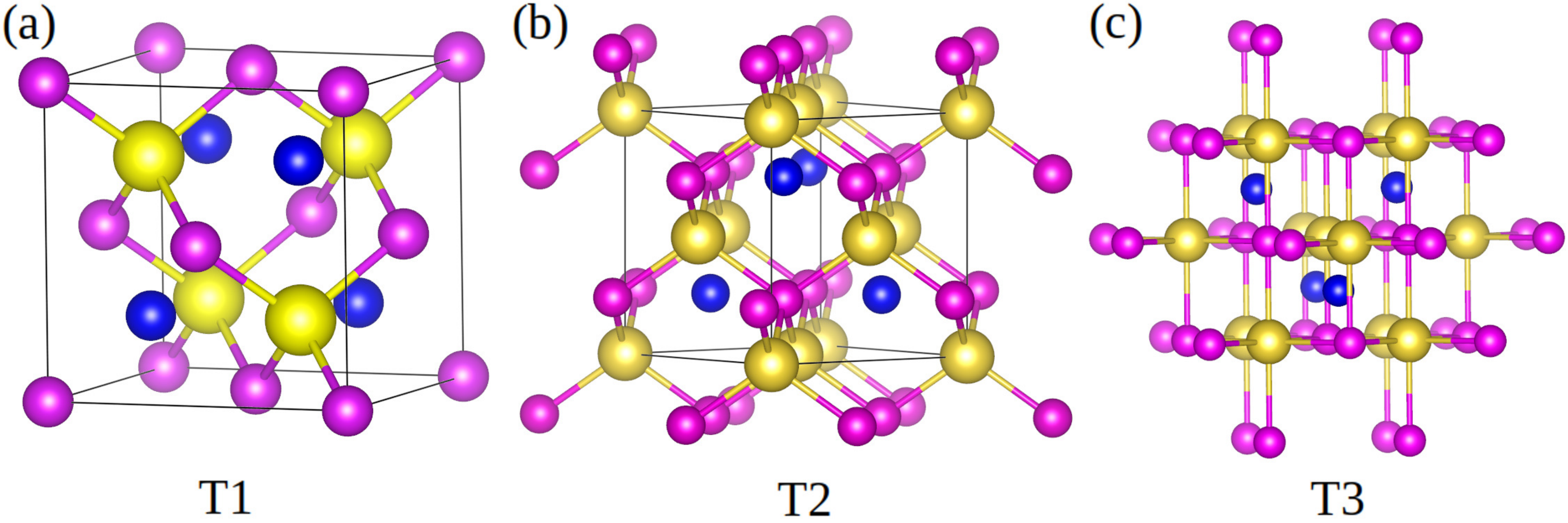}}
\caption{Conventional unit cell of HH $XYZ$ alloys (cubic symmetry, space group $F\overline{4}3m$) considering 
	three possible atomic arrangements. (a)~T1 with Wyckoff positions $4c~(\frac{1}{4},\frac{1}{4},\frac{1}{4})$, 
	$4d~(\frac{3}{4},\frac{3}{4},\frac{3}{4})$, and $4a~(0,0,0)$. (b)~T2 with Wyckoff positions
	$4a~(0,0,0)$, $4d~(\frac{3}{4},\frac{3}{4},\frac{3}{4})$, and $4c~(\frac{1}{4},\frac{1}{4},\frac{1}{4})$.
	(c)~T3 with Wyckoff positions $4b~(\frac{1}{2},\frac{1}{2},\frac{1}{2})$, $4d~(\frac{3}{4},\frac{3}{4},\frac{3}{4})$, 
	and $4a~(0,0,0)$. Yellow, blue, and  pink spheres represent $X$, $Y$, and $Z$ ions, respectively.}
\label{fig1}
\end{figure*}

\section{Methods}
\label{sec:methods}
In what follows, we explain the technical details of our first-principles calculations. We also describe the 
simple spin Heisenberg model that we have devised to analyse the magnetic properties of HH alloys at finite 
temperatures, along with the technical details of the accompanying Monte Carlo simulations. 

\subsection{First-principles calculations}
\label{subsec:DFT}
Density functional theory (DFT) calculations have been performed with the self-consistent full potential 
linearized augmented plane wave (FP-LAPW) method, as implemented in the WIEN2K code \cite{wien2k}. 
The generalised gradient approximation to the exchange-correlation energy due to Perdew-Burke-Ernzerhof
(PBE) \cite{pbe} has been employed in this study to estimate energies and determine equilibrium 
geometries. The value of the R$_{\rm mt} \times$K$_{\rm max}$ product, where R$_{\rm mt}$ is the value of 
the muffin-tin sphere radii and K$_{\rm max}$ of the plane wave cut-off energy, was fixed to $9$ in order to 
provide highly converged results. Likewise, a dense {\bf k}-point mesh of $21 \times 21 \times 21$ was used 
for integrations within the first Brillouin zone (IBZ). The self-consistent threshold values adopted for the 
calculation of energies and equilibrium geometries were $10^{-5}$~eV and $10^{-4}$~eV/\AA, respectively. 
In order to estimate accurate electronic and magnetic properties (e.g., band gaps and magnetic moments) at 
reasonable computational cost, we employed the Tran-Blaha modified Becke-Johnson (TB-mBJ) meta-GGA 
exchange-correlation functional \cite{tran09} over the equilibrium structures generated with PBE.

The pseudopotential plane-wave DFT code VASP \cite{vasp} has been also used for determining the 
energy difference between FM and AFM spin orderings, and for estimating vibrational lattice 
phonons. We used the projector-augmented wave method to represent the ionic cores \cite{bloch94},  
considering the following electrons as valence states: $X$ $s$, $p$, and $d$; $Y$ $s$, $p$, and $d$; 
and $Z$ $s$ and $p$. Wave functions were represented in a plane-wave basis truncated at $650$~eV and
for integrations within the IBZ we employed a dense $\Gamma$-centered {\bf k}-point mesh of $16 \times 
16 \times 16$. The calculation of phonon frequencies was performed with the small displacement method 
\cite{alfe09} and the PHONOPY code \cite{phonopy}. The following parameters provided sufficiently well 
converged phonon frequencies: 190--atom supercells (i.e., $4 \times 4 \times 4$ replication of the 
conventional HH unit cell) and atomic displacements of $0.02$~\AA.

\subsection{Spin Heisenberg model and Monte Carlo simulations}
\label{subsec:spin}
To analyse the effects of temperature on the magnetic properties of HH alloys, we define the following 
spin Heisenberg Hamiltonian:
\begin{equation}
E_{\rm spin} = E_{0} + \frac{1}{2}\sum_{ij} J_{ij}S_{i}S_{j}~, 
\label{eq:heisenberg}
\end{equation}
where $E_{0}$ is a reference energy, $S_{i}$ represent the magnetic moment of ion $i$, and $J_{ij}$ the 
exchange interactions of ion $i$ with the rest. In our model, we only consider spin couplings between 
nearest magnetic ions (which add up to $12$, given that the symmetry of the magnetic sublattice is face 
centered cubic) and assume that all exchange interactions are equal. Based on such simplifications, 
the value of the model parameters can be calculated straightforwardly with DFT methods as:
\begin{eqnarray}
	E_{0} = \frac{1}{2}\left( E_{\rm FM} + E_{\rm AFM} \right) \nonumber \\
	J_{ij} = \frac{1}{8|S|^{2}}\left( E_{\rm FM} - E_{\rm AFM} \right)~,
\label{eq:parameter}
\end{eqnarray}	
where $E_{\rm FM}$ and $E_{\rm AFM}$ are the energy of the crystal considering perfect ferromagnetic and 
anti-ferromagnetic spin arrangements, respectively. (We recall that in the AFM spin configuration each 
magnetic ion sees $8$ out of its $12$ nearest neighbors with opposite spin orientation \cite{singh17}.) It 
is worth noting that in spite of the simplicity of the adopted spin Heisenberg Hamiltonian, similar models 
have been able to provide accurate results for the $T$-dependence of the magnetization of complex magnetic 
materials (e.g., multiferroic oxide perovskites \cite{cazorla13,cazorla17b,cazorla18}).  

Classical Monte Carlo (MC) simulations of the spin Heisenberg Hamiltonian just explained were performed 
to estimate the magnetization of HH alloys (that is, equal to the sum of all invidual spins) as a function 
of temperature. We used a periodically-repeated simulation box containing $20 \times 20 \times 20$ spins. 
Thermal averages were computed from runs of $50,000$ MC sweeps performed after equilibration (which 
consisted of the same number of MC steps). A small symmetry-breaking magnetic anisotropy was introduced 
in the MC simulations to facilitate the accompanying numerical analysis \cite{escorihuela12}. By using this 
setup and monitoring the evolution of the magnetization as a function of $T$, we were able to estimate magnetic 
transition temperatures, $T_{\rm C}$, with a numerical accuracy of $25$~K.

\begin{figure}
\centerline{
\includegraphics[width=1.0\linewidth]{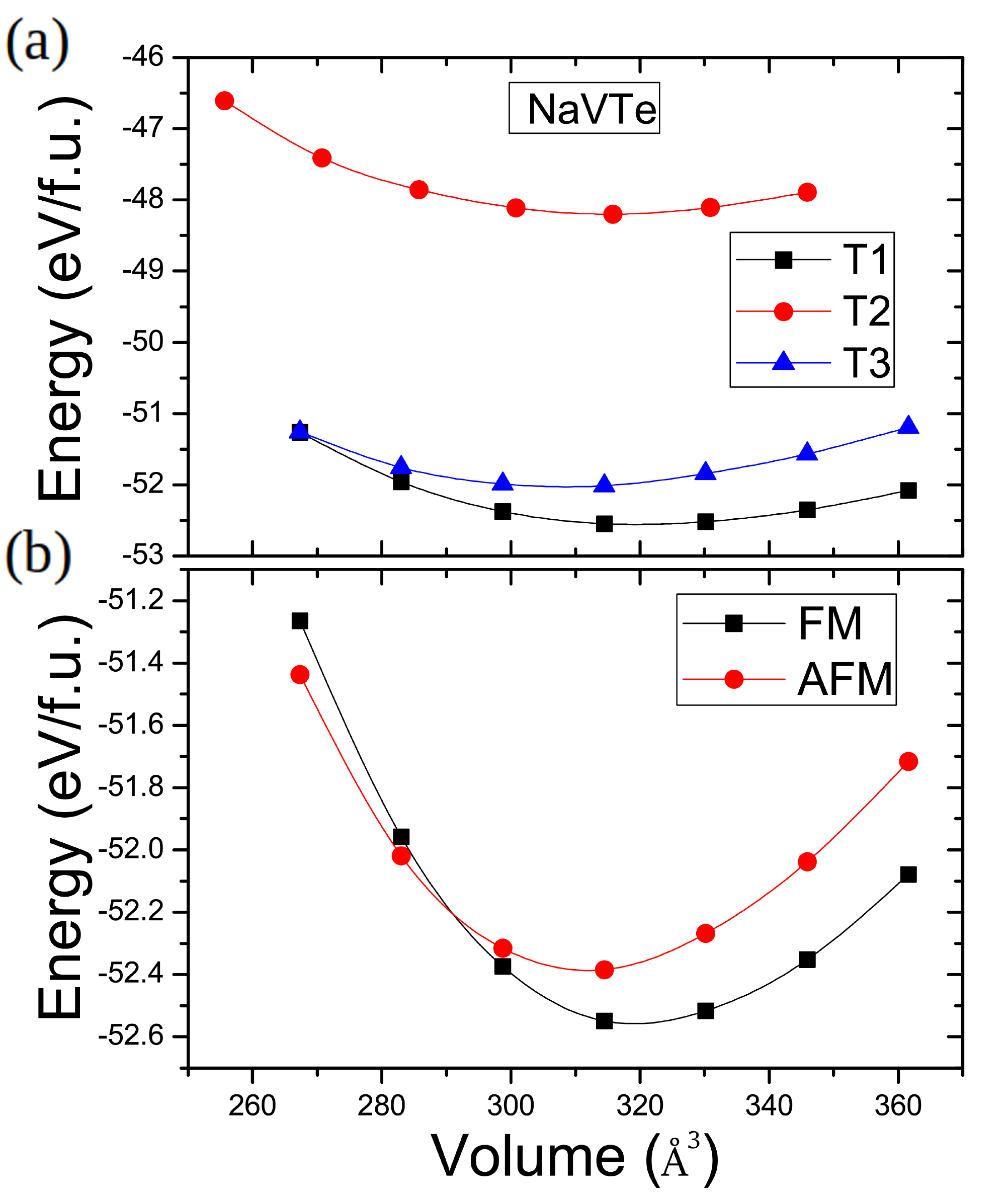}}
\caption{Volume optimization of the HH alloy NaVTe. (a)~Three different atomic structures
	are considered with fixed ferromagnetic (FM) spin ordering. (b)~Ferromagnetic and 
	anti-ferromagnetic (AFM) spin orderings are considered in the lowest-energy T1 structure.}
\label{fig2}
\end{figure}

\begin{figure*}
\centerline{
\includegraphics[width=1.0\linewidth]{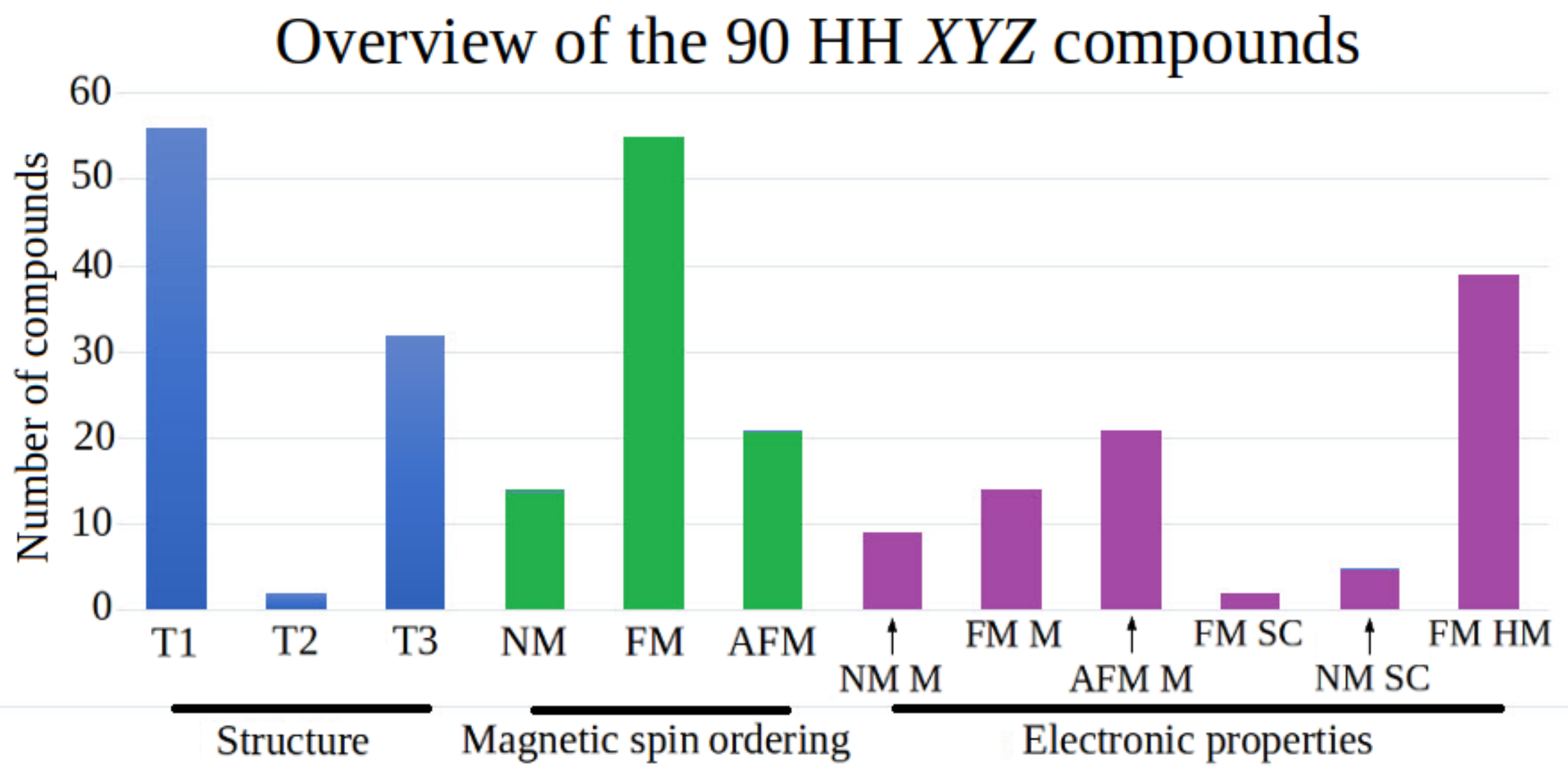}}
\caption{Properties summary of the $90$ HH $XYZ$ alloys analysed in this study for their lowest-energy 
	configuration. Magnetic spin ordering is classified into non-magnetic (NM), ferromagnetic (FM), 
	and anti-ferromagnetic (AFM). Electronic properties are described as metallic (M), semiconductor 
	(SC), and half-metallic (HM).}
\label{fig3}
\end{figure*}

\section{Results and Discussion}
\label{sec:results}
We start this section by providing an overview of the computational strategy that we have followed 
to determine the ground-state configurations of HH alloys, and the general classification that results 
from our calculations based on their stuctural, electronic, and magnetic properties. Then, we explain 
in detail the electronic and vibrational properties of the disclosed HH HM, followed by a discussion 
on their magnetic phase transition temperatures and zero-temperature mixing energies. Finally, we 
highlight the HH alloys that according to our calculations are most promising for spintronics
applications. 

\begin{figure*}
\centerline{
\includegraphics[width=1.0\linewidth]{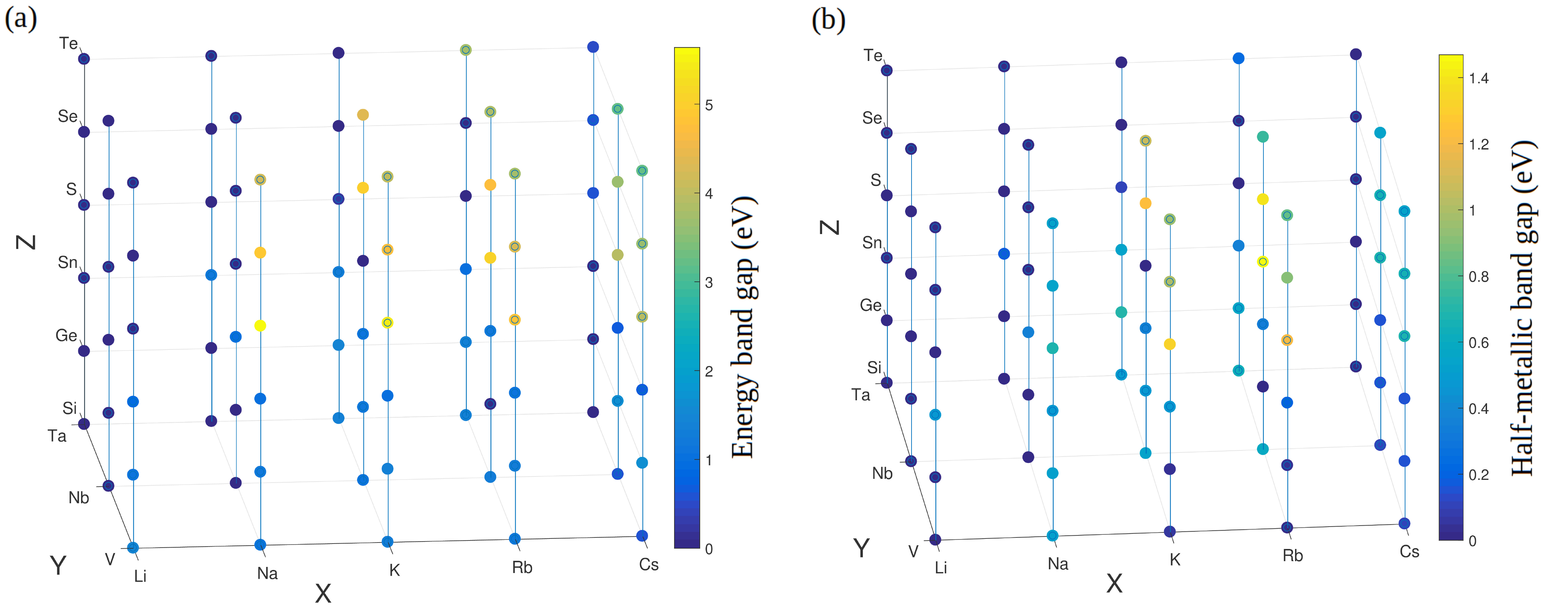}}
\caption{Electronic properties of the $90$ HH $XYZ$ alloys analysed in this study for their lowest-energy
	configuration. (a)~Energy band gap, $E_{\rm BG}$, and (b)~half-metallic band gap, $E_{\rm HM}$. Blue
	and yellow colors denote smaller and larger band gap values, respectively, and the integrating $X$, 
	$Y$, and $Z$ species are indicated in the three orthogonal axis.}
\label{fig4}
\end{figure*}

\begin{figure*}
\centerline{
\includegraphics[width=1.0\linewidth]{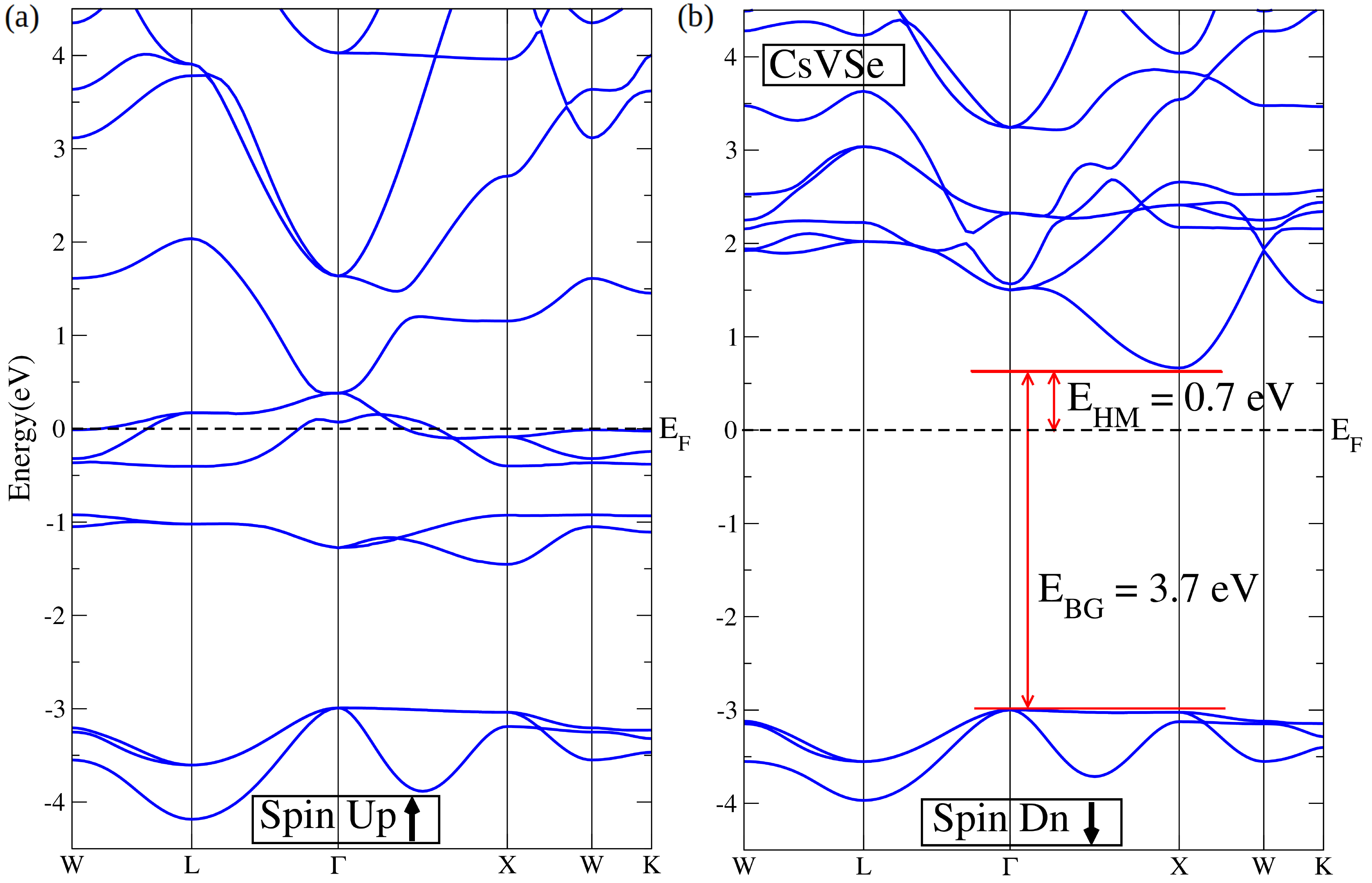}}
\caption{Band structure of the half-metallic HH alloy CsVSe (T3 arrangement) predicted in this
         study; the horizontal dashed line represents the Fermi level, $E_{\rm F}$, which has
         been shifted to zero. CsVSe is vibrationally stable at $\Gamma$, displays FM spin
         ordering with a Curie temperature of $T_{\rm C} = 375 \pm 25$~K, possesses a
         relatively small zero-temperature mixing energy of $0.11$~eV/atom, and has a
         total magnetic moment of $4$~$\mu_{\rm B}$. The energy band gap of CsVSe is
         $3.66$~eV and the accompanying half-metallic band gap is $0.66$~eV. The electronic
         properties are calculated with the Tran-Blaha modified Becke-Johnson (TB-mBJ)
         meta-GGA exchange-correlation functional \cite{tran09}.}
\label{fig5}
\end{figure*}

\subsection{Overview of the $90$ $XYZ$ HH alloys}
\label{subsec:overview}
The $XYZ$ HH alloys considered in this study feature an alkali metal in the $X$ position (Li, Na, K, Rb, 
and Cs), a transition metal in the $Y$ position (V, Nb, and Ta), and a non-metal $sp$-element in the $Z$ 
position (Si, Ge, Sn, S, Se, and Te). The cubic HH unit cell consists of three interpenetrating face centered 
cubic (fcc) sublattices that together render a crystal structure with $F\overline{4}3m$ symmetry. There exist 
three possible atomic arrangements compatible with this structure, namely, T1, T2, and T3, which are generated 
by exchanging the Wyckoff positions of the $X$, $Y$, and $Z$ ions (see Fig.~\ref{fig1}). The physical properties 
of HH alloys may be greatly influenced by their specific atomic arrangement \cite{sattar18}, hence configurations 
T1, T2, and T3 need to be all considered when performing thorough functionality searches within this family
of materials.

Systematic fixed-volume geometry optimizations have been carried out for each HH compound to determine the 
energetically most favorable atomic arrangement, equilibrium volume, and magnetic ordering. First, the equilibrium volume 
of each possible structure considering FM spin ordering is obtained by fitting the series of calculated energy points  
to a Birch-Murnaghan equation of state \cite{cazorla15}; subsequently, the energy associated with AFM spin ordering 
is analyzed for the lowest-energy atomic arrangement obtained in the FM case. Figures~\ref{fig2}a-b illustrate such 
a computational procedure for the particular case of NaVTe, which turns out to exhibit a T1--FM ground state and an 
equilibrium volume of $320$~\AA$^{3}$. Once the optimal atomic arrangement, equilibrium volume, and magnetic ordering 
have been determined at the PBE level, we accurately calculate the corresponding electronic properties (e.g., energy 
band gaps and magnetic moments) with the TB-mBJ functional \cite{tran09}.

Figure~\ref{fig3} shows a general classification of the $90$ HH compounds analyzed in this study made on basis 
to their structural, magnetic, and electronic properties. A total of $56$ materials present lowest energy on the
T1 arrangement, $32$ on the T3, and only $2$ on the T2. Compounds with a light-weight alkali metal in the $X$ 
position (e.g., Li, Na, and K) tend to be more stable in the T1 arrangement, whereas those with heavy-weight 
alkali metals (e.g., Rb and Cs) in the T3. On the other hand, compounds with a heavy-weight transition metal
in the $Y$ position (e.g., Nb and Ta) mostly are stabilized in the T1 configuration, made the exception of 
the HH alloys containing Cs which always adopt the T3 arrangement. Meanwhile, the role of the non-metal element 
occupying the $Z$ position on choosing the most favorable structure appears to be negligible. We note that the 
two compounds that adopt the T2 configuration turn out to be metallic (i.e., LiTaSn and LiNbSn); thereby, the 
T2 arrangement will be ignored for the remainder of the article.  

At zero-temperature conditions, we find that $55$ HH alloys are FM, $21$ AFM, and $14$ non-magnetic (see 
Fig.\ref{fig3}). A total of $39$ FM compounds are found to be half-metallic, $2$ FM semiconductor (SC), and 
the rest metallic. In what follows, we describe with detail the electronic, vibrational, magnetic, and mixing 
properties of the $41$ new HM and SC compounds that have been determined in our investigation, all of which 
present FM spin ordering. 

\begin{figure}
\centerline{
\includegraphics[width=1.0\linewidth]{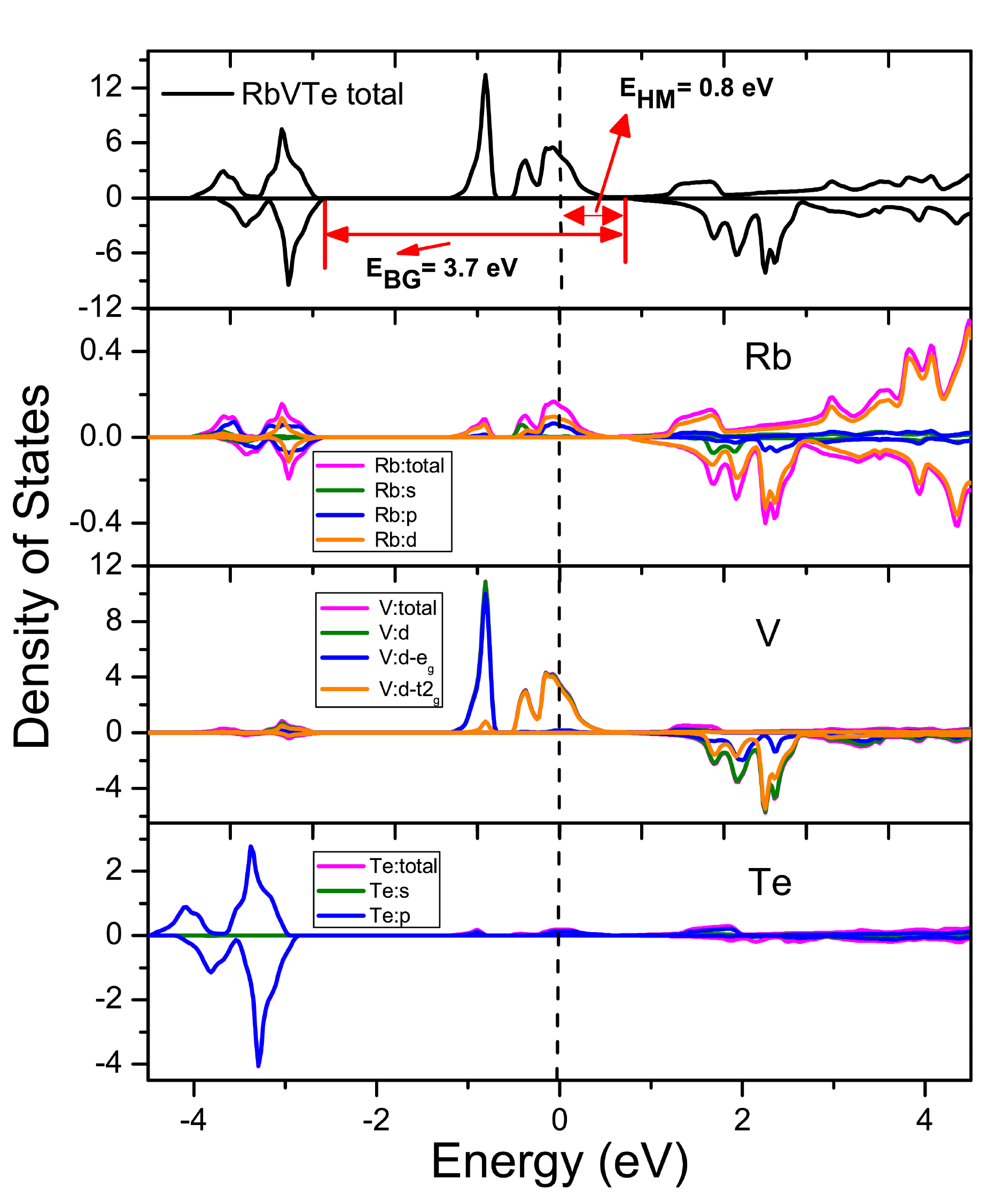}}
\caption{Spin-projected total and partial density of states of the half-metallic HH alloy
         RbVTe (T3 arrangement) predicted in this study; spin up~(down) density components are
         represented in the positive~(negative) panels. The horizontal dashed line represents
         the Fermi level, which has been shifted to zero. RbVTe is vibrationally stable at $\Gamma$,
         displays FM spin ordering with a Curie temperature of $T_{\rm C} = 375 \pm 25$~K, possesses
         a relatively small zero-temperature mixing energy of $0.14$~eV/atom, and has a total
         magnetic moment of $4$~$\mu_{\rm B}$. The energy band gap of RbVTe is $3.73$~eV and the
         accompanying half-metallic band gap is $0.83$~eV. The electronic properties are calculated
         with the Tran-Blaha modified Becke-Johnson (TB-mBJ) meta-GGA exchange-correlation
         functional \cite{tran09}.}
\label{fig6}
\end{figure}

\subsection{Electronic and vibrational properties}
\label{subsec:zerotemp}
Figure~\ref{fig4} shows the energy and half-metallic band gaps, $E_{\rm BG}$ and $E_{\rm HM}$, estimated for 
the $90$ $XYZ$ HH alloys considered in this study. $E_{\rm BG}$ is the energy difference between the top of 
the valence band and bottom of the conduction band in the semiconductor spin channel, and $E_{\rm HM}$ the 
minimum energy that an electron requires to surpass the spin gap (see Fig.\ref{fig5}, where we represent 
those quantities for CsVSe). One can observe a certain correlation between these two quantities, namely, 
large $E_{\rm BG}$'s are accompanied by large $E_{\rm HM}$'s. For instance, RbNbS (T1 arrangement) 
has an energy band gap of $5.08$~eV and HM band gap of $1.46$~eV, while CsVSn (T3 arrangement) has $1.30$~eV 
and $0.18$~eV, respectively. Nevertheless, there are few compounds that in spite of having a relatively small 
energy band gap exhibit a relatively large HM band gap (e.g., RbNbSi in the T3 configuration for which we 
estimate $E_{\rm BG} = 1.24$~eV and $E_{\rm HM} = 0.57$~eV). 

The color code employed in Fig.\ref{fig4} shows that the HH alloys consisting of heavy-weight $X$-ions (K, 
Rb, and Cs), light-weight $Y$-ions (V and Nb), and chalcogen elements (S, Se, and Te) in the $Z$ position, 
generally possess the largest $E_{\rm BG}$ and $E_{\rm HM}$. By contrast, HH compounds with a light-weight~(heavy-weight) 
alkali~(transition) metal element like Li~(Ta) in the $X$~($Y$) position mostly are metallic or present 
medium band gaps.

Figure~\ref{fig6} shows the spin-projected density of electronic states calculated for the 
ground state of RbVTe (T3 configuration). This compound is half-metallic with a large 
$E_{\rm HM}$ of $0.83$~eV and large energy band gap of $3.73$~eV, hence is an interesting 
case for which to analyze the electronic features. In the majority band (spin up), 
is appreciated that the V $d$--$t_{2g}$ and Rb $d$ orbitals are highly hybridized in the region 
surrounding the Fermi level, which leads to the appearance of metallicity in that channel. In 
the minority band (spin down), however, the V and Rb $d$ orbitals are shifted to higher 
energies and remain unoccupied; consequently, a HM band gap appears in that channel between 
the occupied $d$ bonding and unoccupied $d$ antibonding states. Meanwhile, the majority of 
states forming the top of the valence band in the minority band (spin down) correspond to Te $p$ 
orbitals, which are shifted to energies well below those of the V and Rb $d$ orbitals. 

We note that when the chalcogen element (S, Se, and Te) in the $Z$ position is substituted by a 
group--XIV element (Si, Ge, Sn), $E_{\rm BG}$ and $E_{\rm HM}$ generally undergo a significant 
reduction (see Fig.\ref{fig4}). For instance, RbVSn has an energy band gap of $1.13$~eV and 
a HM band gap of $0.20$~eV. Such a band gap cloisure is consequence of a decrease in the energy 
separation between the $Z$ $p$ orbitals and $X$--$Y$ $d$ bonding and antibonding states (not shown
here), and it occurs when the number of valence electrons in the system changes from $12$ (S, Se, 
Te) to $10$ (Si, Ge, Sn). Interestingly, the energy band gap changes as induced by $Z$--element 
substitutions correlate directly with changes in the magnetic moment of the ions; we will discuss 
in detail this and other magnetic effects in the next subsection. 

\begin{figure*}
\centerline{
\includegraphics[width=1.0\linewidth]{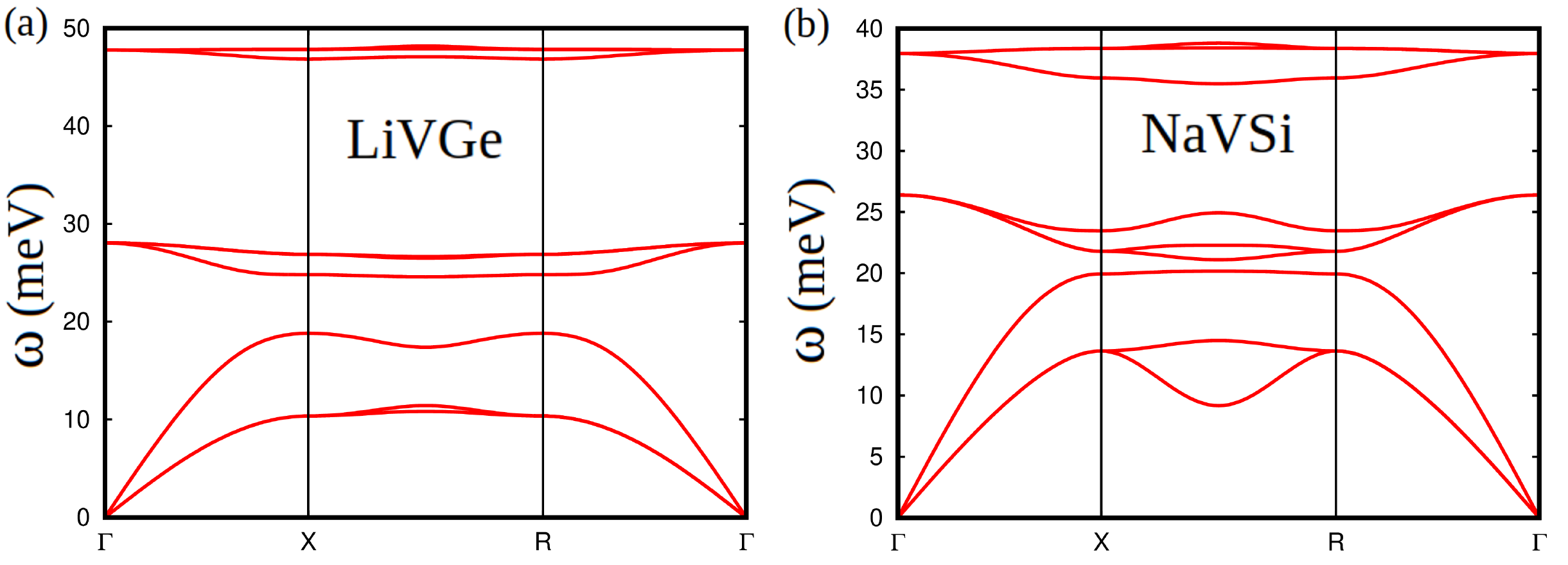}}
\caption{Phonon spectrum of the semiconductor and half-metallic HH alloys LiVGe and NaVSi
	 (both in the T1 arrangement) predicted in this study. (a)~LiVGe is vibrationally stable, 
	 displays FM spin ordering with a Curie temperature of $T_{\rm C} = 375 \pm 25$~K, 
	 possesses a zero-temperature mixing energy of $-0.11$~eV/atom, a total magnetic moment 
	 of $2$~$\mu_{\rm B}$, and an energy band gap of $1.06$~eV. (b)~NaVSi is vibrationally 
	 stable, displays FM spin ordering with a Curie temperature of $T_{\rm C} = 300 \pm 25$~K, 
	 possesses a relatively small zero-temperature mixing energy of $0.21$~eV/atom, a total 
	 magnetic moment of $2$~$\mu_{\rm B}$, an energy band gap of $1.11$~eV, and an accompanying 
	 half-metallic band gap of $0.52$~eV.}
\label{fig7}
\end{figure*}

Out of the $90$ HH alloys investigated in this study, we have found that $39$~($2$) are ferromagnetic
and half-metallic~(magnetic semiconductor) at zero-temperature conditions. In order to provide useful guides 
for the experiments, is necessary also to assess the vibrational stability of the candidate materials 
proposed by theory. Here, we have calculated the lattice phonons for each HH alloy at the high-symmetry 
reciprocal point $\Gamma$, and concluded that $28$ FM HM along with the $2$ FM SC are vibrationally well 
behaved (that is, do not present imaginary phonon frequencies at the IBZ center). For some selected cases 
(namely, the overall most promising materials that will be discussed in Sec.\ref{subsec:promise}), we have 
calculated also the full phonon spectrum over the entire IBZ and found that most of them are vibrationally 
stable (see Fig.\ref{fig7}, where we enclose a couple of relevant examples). We have not been able to draw
any robust correlation between vibrational instability and chemical structure for the HH alloys analyzed
in this study. For the remainder of the article, we will focus on describing the magnetic and mixing 
properties of the $28$~($2$) FM, HM~(SC), and vibrational stable compounds that we have predicted. 

\begin{figure}
\centerline{
\includegraphics[width=1.0\linewidth]{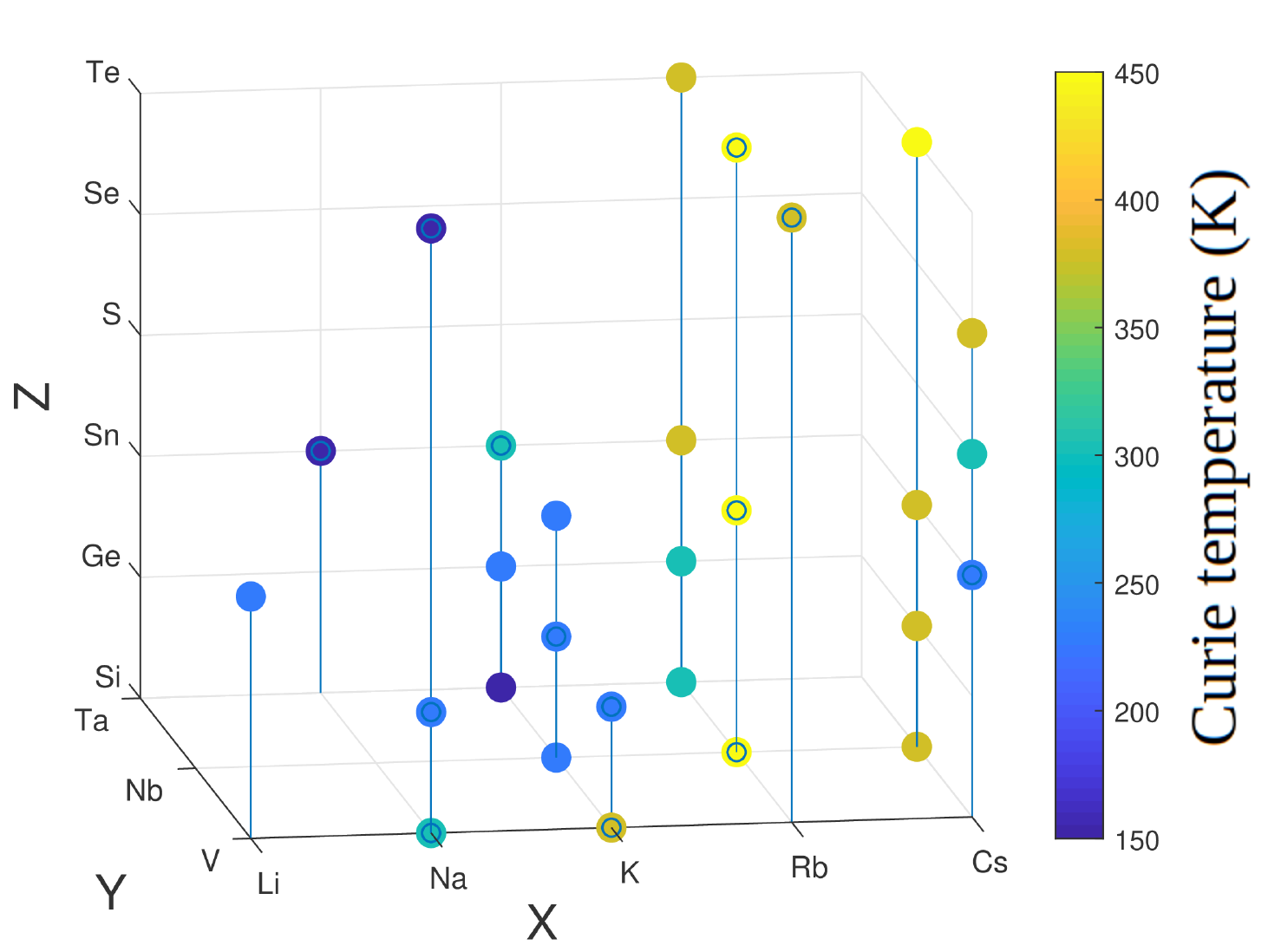}}
\caption{Curie temperature, $T_{\rm C}$, of the $28$ half-metallic HH alloys found in this study 
	to be vibrationally stable at $\Gamma$ and which display FM spin ordering. Yellowish color 
	indicates magnetic transition points above room temperature; the numerical uncertainty in 
	our $T_{\rm C}$ results is $\pm 25$~K.}
\label{fig8}
\end{figure}

\begin{figure}
\centerline{
\includegraphics[width=1.0\linewidth]{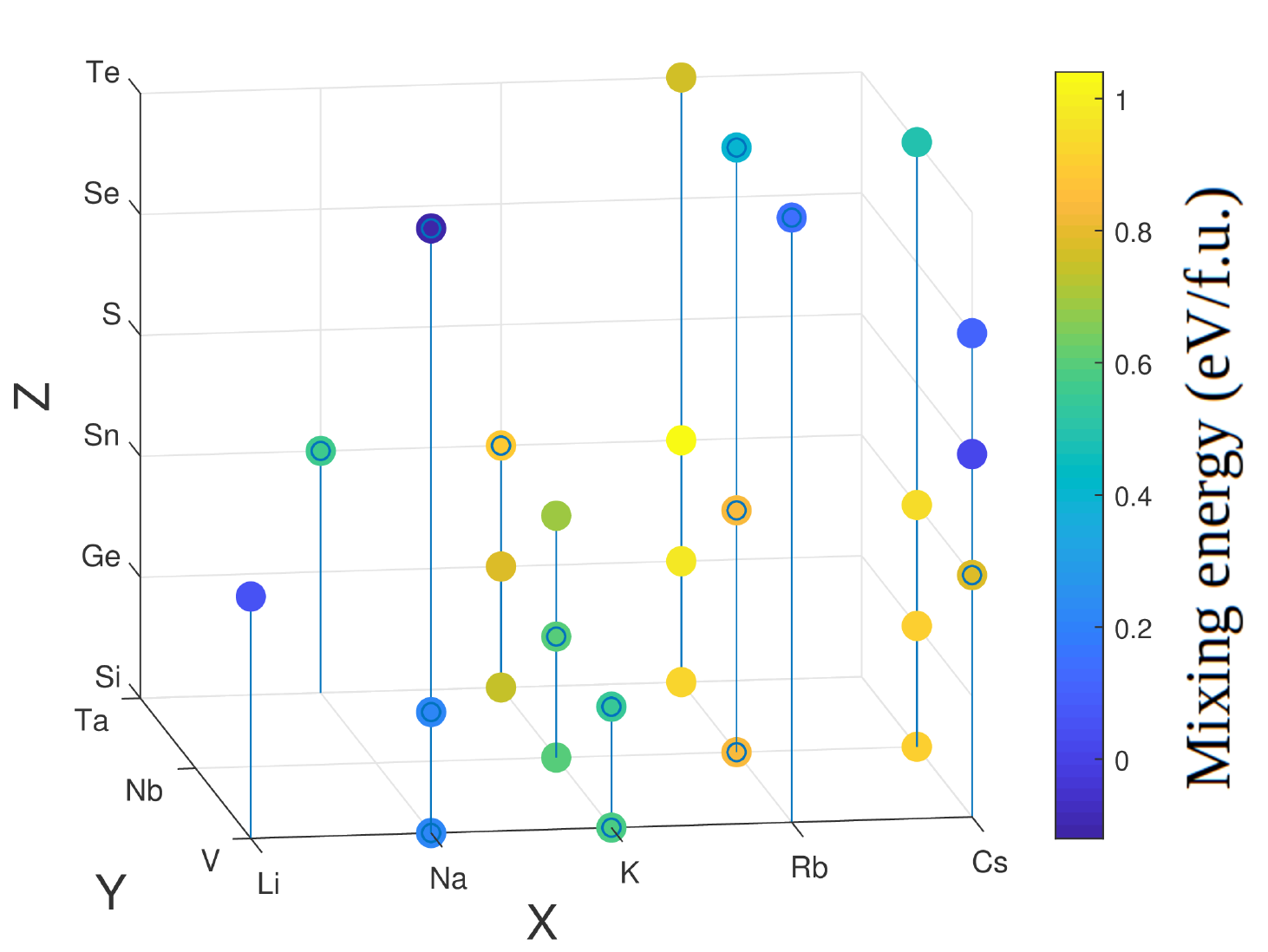}}
\caption{Zero-temperature mixing energy, $E_{\rm mix}$, of the $28$ half-metallic HH alloys found
        in this study to be vibrationally stable at $\Gamma$ and which display FM spin ordering.
        Bluish color indicates mixing energy values below $0.2$~eV per formula unit.}
\label{fig9}
\end{figure}

\begin{table*}
\centering
\begin{tabular}{c c c c c c c c c}
\hline
\hline
$ $ & $ $ & $ $ & $ $ & $ $ & $ $ & $ $ & $ $ & $ $ \\
	$ \quad {\rm Compound} \quad $ & $ \quad {\rm Structure} \quad $ & $ \quad a_{0} \quad $ & $ \quad {\rm Electronic} \quad $ & $ \quad M \quad $ & $ \quad E_{\rm BG} \quad $ & $ \quad E_{\rm HM} \quad $ & $ \quad T_{\rm C} \quad $ & $ \quad E_{\rm mix} \quad $ \\
	$ $ & $ $ & ${\rm (\AA)}$ & $ {\rm behavior} $ & $ (\mu_{\rm B}) $ & $ {\rm (eV)} $ & $ {\rm (eV)} $ & $ {\rm (K)} $ & $ ({\rm eV/f.u.}) $ \\
$ $ & $ $ & $ $ & $ $ & $ $ & $ $ & $ $ \\
\hline
$ $ & $ $ & $ $ & $ $ & $ $ & $ $ & $ $ & $ $ & $ $ \\
	${\bf LiVSi}  $ & ${\rm T1} $ & $ 5.85 $ & $ {\bf SC} $ & $ {\bf 2.0} $ & $ {\bf 1.43} $ & $ -    $ & $ {\bf 300} $ & $ {\bf -0.11} $ \\
	${\bf LiVGe}  $ & ${\rm T1} $ & $ 5.93 $ & $ {\bf SC} $ & $ {\bf 2.0} $ & $ {\bf 1.06} $ & $ -    $ & $ {\bf 375} $ & $ {\bf -0.11} $ \\
$ $ & $ $ & $ $ & $ $ & $ $ & $ $ & $ $ & $ $ & $ $ \\
	${\bf NaVSi}  $ & ${\rm T1} $ & $ 6.25 $ & $ {\bf HM} $ & $ {\bf 2.0} $ & $ {\bf 1.11} $ & $ {\bf 0.52} $ & $ {\bf 300} $ & $ {\bf +0.21} $ \\
	${\rm NaVTe}  $ & ${\rm T1} $ & $ 6.84 $ & $ {\rm HM} $ & $ 4.0 $ & $ 4.18 $ & $ 0.54 $ & $ 150 $ & $ -0.12 $ \\
	${\rm KVSi}   $ & ${\rm T3} $ & $ 6.55 $ & $ {\rm HM} $ & $ 2.0 $ & $ 1.49 $ & $ 0.21 $ & $ 375 $ & $ +0.56 $ \\
	${\bf RbVTe}  $ & ${\rm T3} $ & $ 7.27 $ & $ {\bf HM} $ & $ {\bf 4.0} $ & $ {\bf 3.73} $ & $ {\bf 0.83} $ & $ {\bf 375} $ & $ {\bf +0.14} $ \\
	${\bf CsVS}   $ & ${\rm T3} $ & $ 6.93 $ & $ {\bf HM} $ & $ {\bf 4.0} $ & $ {\bf 3.92} $ & $ {\bf 0.68} $ & $ {\bf 300} $ & $ {\bf +0.01} $ \\
	${\bf CsVSe}  $ & ${\rm T3} $ & $ 7.13 $ & $ {\bf HM} $ & $ {\bf 4.0} $ & $ {\bf 3.66} $ & $ {\bf 0.66} $ & $ {\bf 375} $ & $ {\bf +0.11} $ \\
$ $ & $ $ & $ $ & $ $ & $ $ & $ $ & $ $ \\
	${\rm RbNbSi} $ & ${\rm T3} $ & $ 6.77 $ & $ {\rm HM} $ & $ 2.0 $ & $ 1.24 $ & $ 0.57 $ & $ 450 $ & $ +0.83 $ \\
	${\rm RbNbSn} $ & ${\rm T3} $ & $ 7.19 $ & $ {\rm HM} $ & $ 2.0 $ & $ 1.19 $ & $ 0.33 $ & $ 450 $ & $ +0.83 $ \\
	${\bf RbNbTe} $ & ${\rm T3} $ & $ 7.41 $ & $ {\bf HM} $ & $ {\bf 4.0} $ & $ {\bf 3.82} $ & $ {\bf 0.75} $ & $ {\bf 450} $ & $ {\bf +0.40} $ \\
	${\rm CsNbSi} $ & ${\rm T3} $ & $ 6.87 $ & $ {\rm HM} $ & $ 2.0 $ & $ 0.47 $ & $ 0.11 $ & $ 375 $ & $ +0.91 $ \\
	${\rm CsNbGe} $ & ${\rm T3} $ & $ 6.96 $ & $ {\rm HM} $ & $ 2.0 $ & $ 0.62 $ & $ 0.14 $ & $ 375 $ & $ +0.91 $ \\
	${\rm CsNbSn} $ & ${\rm T3} $ & $ 7.28 $ & $ {\rm HM} $ & $ 2.0 $ & $ 0.69 $ & $ 0.16 $ & $ 375 $ & $ +0.94 $ \\
	${\rm CsNbTe} $ & ${\rm T3} $ & $ 7.55 $ & $ {\rm HM} $ & $ 4.0 $ & $ 3.23 $ & $ 0.39 $ & $ 450 $ & $ +0.49 $ \\
$ $ & $ $ & $ $ & $ $ & $ $ & $ $ & $ $ \\
	${\rm KTaSn}  $ & ${\rm T1} $ & $ 7.17 $ & $ {\rm HM} $ & $ 2.0 $ & $ 1.31 $ & $ 0.54 $ & $ 300 $ & $ +0.88 $ \\
	${\rm RbTaSi} $ & ${\rm T1} $ & $ 6.89 $ & $ {\rm HM} $ & $ 2.0 $ & $ 1.36 $ & $ 0.66 $ & $ 300 $ & $ +0.92 $ \\
	${\rm RbTaGe} $ & ${\rm T1} $ & $ 6.99 $ & $ {\rm HM} $ & $ 2.0 $ & $ 1.39 $ & $ 0.55 $ & $ 300 $ & $ +0.97 $ \\
	${\rm RbTaSn} $ & ${\rm T3} $ & $ 7.16 $ & $ {\rm HM} $ & $ 2.0 $ & $ 1.05 $ & $ 0.35 $ & $ 375 $ & $ +1.03 $ \\
	${\rm RbTaTe} $ & ${\rm T3} $ & $ 7.37 $ & $ {\rm HM} $ & $ 4.0 $ & $ 3.70 $ & $ 0.24 $ & $ 375 $ & $ +0.75 $ \\
$ $ & $ $ & $ $ & $ $ & $ $ & $ $ & $ $ & $ $ & $ $ \\
\hline
\hline
\end{tabular}
\label{tab:summary}
\caption{Properties summary of the most relevant HH alloys predicted in this study considering their lowest-energy configuration. 
	The $7$ overall most promising HH compounds are highlighted in bold. All systems are vibrationally stable and display 
	FM spin ordering. The electronic properties are calculated with the Tran-Blaha modified Becke-Johnson (TB-mBJ) meta-GGA 
	exchange-correlation functional \cite{tran09}. The numerical uncertainty in our $T_{\rm C}$ results is $\pm 25$~K.}
\end{table*}

\subsection{Magnetism and mixing stability}
\label{subsec:finitetemp}
Recently, Damewood \emph{et al.} \cite{damewood15} have proposed a modified Slater-Pauling (mSP)
rule for describing the magnetic moment of the half-Heusler alloys LiMn$Z$ ($Z =$ N, P, Si).
Specifically, the total magnetic moment per formula unit (expressed in units of $\mu_{\rm B}$), $M$,
has been parametrized as:
\begin{equation}
M = N_{t} - 8~,
\label{eq:msp}
\end{equation}
where $N_{t}$ is the total number of valence electrons in the unit cell. In the present case,
$N_{t}$ is equal to $12$ for alloys displaying S, Se, or Te in the $Z$ position, and to $10$ for Si,
Ge, and Sn (we recall that $N_{t} = 1$ for the $X$ elements considered in this study, $N_{t} = 5$
for the $Y$'s, and $N_{t} = 4$ or $6$ for the $Z$'s). We have found that all non-metallic and
magnetic $XYZ$ alloys disclosed in this study (i.e., either HM or SC) in fact follow the mSP
rule proposed by Damewood \emph{et al.} \cite{damewood15} (some examples are provided in Table~I).
Interestingly, in those cases we also observe a clear correlation between the total magnetization
and size of the energy band gap: large $M$'s and large $E_{\rm BG}$'s occur simultaneously 
in the same compounds. From an applied point of view, this seems to be a positive finding 
since wide band gap HM alloys then are likely to be also magnetically robust. Nevertheless, 
the size of the magnetic moments does not provide any information on the variation of the
magnetization at $T \neq 0$ conditions, which is related to the exchange interactions between
the magnetic moments. 

We have estimated the magnetic transition temperature, $T_{\rm C}$, for the $28$~($2$) FM, HM~(SC),
and vibrational stable compounds determined in this study, by using the computational methods
explained in Sec.~\ref{subsec:spin}. Our $T_{\rm C}$ results are shown in Fig.\ref{fig8}. It is
appreciated that compounds with a heavy-weight alkali metal in the $X$ position (Rb and Cs)
concentrate the highest magnetic transition temperatures (see also Table~I). For instance, RbNbTe
displays a Curie temperature of $450 \pm 25$~K and CsVSe of $375 \pm 25$~K, while NaVTe becomes
paramagnetic at $150 \pm 25$~K and KTaGe at $225 \pm 25$~K. We ascertain the lack of any
correlation between the size of the ionic magnetic moments, $M$, and the Curie temperature of 
the system; for example, RbNbSn displays $M = 2~\mu_{\rm B}$ and $T_{\rm C} = 450 \pm 25$~K while 
for RbNbTe we estimate $M = 4~\mu_{\rm B}$ and the same magnetic transition temperature. Interestingly, 
we find that a total of $17$~($2$) FM, HM~(SC), and vibrational stable compounds remain magnetically 
ordered at or above room temperature. All of them are listed in Table~I and will be discussed in 
the next subsection.

Finally, we calculated the zero-temperature mixing energy, $E_{\rm mix}$, of the $28$~($2$) FM, HM~(SC), 
and vibrational stable compounds disclosed in this work, by using the formula:
\begin{equation}
	E_{\rm mix} = E_{XYZ}^{\rm HH} - \left( E_{X}^{\rm fcc} + E_{Y}^{\rm fcc} + E_{Z}^{\rm fcc} \right)~,
\label{eq:mix}
\end{equation}	
where $E_{XYZ}^{\rm HH}$ represents the ground-state energy of the HH alloy, and $E_{A}^{\rm fcc}$ the energy 
of the bulk $A$ crystal considering an equilibrium fcc structure. $E_{\rm mix}$ provides 
a quantitative estimation of how stable the $XYZ$ system is against decomposition into $X$-, $Y$-, and $Z$-rich 
regions. In particular, negative~(positive) values of the mixing energy indicate high~(low) stability of the compound 
against phase decomposition. We should note, however, that the configurational entropy of the HH alloy, which 
is totally neglected in Eq.(\ref{eq:mix}), will always contribute favourably to the free energy and enhance the 
chemical stability of the $XYZ$ compound at finite temperatures \cite{page16,rost15}. Consequently, a positive 
but small value of $E_{\rm mix}$ does not necessarily imply phase separation under realistic $T \neq 0$~K conditions. 
Here, we (somewhat arbitrarily) consider that a $E_{\rm mix}$ threshold value of $0.2$~eV per formula unit can be used to 
sieve materials with reasonably good mixing stability from those with tendency for phase separation \cite{shenoy19}. 

Figure~\ref{fig9} shows our $E_{\rm mix}$ results; only $7$ compounds out of $30$ (i.e., $28$ HM and $2$ SC) 
display zero-temperature mixing energies below $0.2$~eV per formula unit. We note that all of those low mixing 
energy compounds contain V in the $Y$ position, namely, LiVSi (SC), LiVGe (SC), NaVTe (HM), NaVSi (HM), RbVTe 
(HM), CsVS (HM), and CsVSe (HM) (see Table~I). Specifically, only the first three alloys listed above present 
negative $E_{\rm mix}$ values, while for CsVS we obtain a practically null mixing energy. On the other hand, KVSi 
(HM), RbNbTe (HM), and CsNbTe (HM) exhibit mixing energies close to $0.50$~eV/f.u., and for the rest of compounds 
we estimate $E_{\rm mix}$'s that are close to $1.00$~eV per formula unit. These results indicate that most 
of the HH HM reported in this work, in spite of possessing relatively high Curie temperatures, are likely to 
present phase separation issues, which is not desirable for practical applications.

Meanwhile, we observe that when the element occupying the $Z$ position in the HH alloy is a chalcogen (S, Se, 
and Te) the resulting mixing energy is noticeably smaller than when the element belongs to group--XIV of the
periodic table (Si, Ge, Sn). For instance, we find that $E_{\rm mix}$ amounts to $-0.12$~eV/f.u. for NaVTe 
and to $0.21$~eV/f.u. for NaVSi (see Table~I). The same behaviour is observed also for other compounds presenting 
either Nb or Ta in the $Y$ position (e.g., for RbTaSn we estimate $1.03$~eV/f.u. and for RbTaTe $0.75$~eV/f.u.). 
Therefore, a possible strategy for improving the mixing stability of some of the new HH HM compounds reported 
in this study (e.g., KVSi with $E_{\rm mix} = 0.56$~eV/f.u. and RbNbSi with $E_{\rm mix} = 0.83$~eV/f.u.) may 
consist in doping with light-weight alkali metals (Li and Na) in the $X$ position (although this may also lead
to some unwanted decrease in the energy band gap and Curie temperature of the alloy, see present and previous 
sections) and with chalcogen species in the $Z$ position.

\subsection{Most promising magnetic HH alloys}
\label{subsec:promise}
Table~I shows the $19$ vibrationally stable and ferromagnetic compounds predicted in this study that possess 
a Curie temperature equal or above room temperature. Two out of those $19$ alloys are semiconductor while the 
rest are half-metallic. We have also included NaVTe in the table, in spite of presenting a relatively low Curie 
temperature of $150$~K, owing to its good mixing stability properties ($E_{\rm mix} = -0.12$~eV/f.u.). From 
an applied perspective, an overall promising HM (or magnetic semiconductor) material should present the following 
qualities: (1)~being vibrationally and chemically stable, (2)~high Curie temperature, (3)~large energy band gaps, 
(4)~large magnetic moment, and (5)~being structurally compatible with other semiconductor materials typically 
employed in electronic devices (e.g., silicon and GaAs with respective lattice parameters of $5.4$ and 
$5.6$~\AA).  

Except NaVTe and CsNb$Z$ with $Z =$ Si, Ge, and Sn, all the compounds reported in Table~I fulfill conditions 
(2), (3), and (4) above. Besides, compounds LiVSi (SC), LiVGe (SC), and NaVSi (HM) also fulfill (1) and 
(5), which indicates that these materials are in fact very promising for spintronics applications \cite{tu16,cao17}. 
Meanwhile, RbVTe (HM), CsVS (HM), and CsVSe (HM) fulfill (1)--(4) and only partially (5); however, the energy 
band gaps estimated for these compounds are so large that they also deserve to be highlighted. Finally, we 
mention RbNbTe as the most encouraging case of a HH HM not containing vanadium in the $Y$ position; this alloy 
fulfills conditions (2)--(4), as many other compounds, but the corresponding $E_{\rm BG}$, $E_{\rm HM}$, and 
$T_{\rm C}$ values are exceedingly large. 

Overall, the most promising HH alloys for use in spintronics applications predicted by our computational 
research are LiVSi (SC), LiVGe (SC), NaVSi (HM), RbVTe (HM), CsVS (HM), CsVSe (HM) and RbNbTe (HM), 
all of which possess magnetic transition temperatures at or above room temperature. The only apparent 
disadvantage of some of these compounds (i.e., NaVSi and RbNbTe) are their positive and large mixing 
energies (see Table~I), which suggests the likely existence of phase separation issues in practice. 
Nevertheless, as we have mentioned earlier, a likely stategy for solving this problem may consist in doping 
to some extent with light-weight alkali metals (Li and Na) in the $X$ position, which in turn would improve 
their structural compatibility with typical semiconductor materials, and/or with chalcogen atoms in the 
$Z$ position.

\section{Conclusions}
\label{sec:conclusions}
We have performed a comprehensive first-principles study of the structural, electronic, structural, vibrational, 
and mixing properties of $90$ $XYZ$ half-Heusler alloys ($X =$~Li, Na, K, Rb, Cs; $Y =$~V, Nb, Ta; $Z =$~Si, 
Ge, Sn, S, Se, Te). In contrast to previous computational studies dealing with a large number of candidate 
materials, we have analyzed the magnetic features of most HH alloys at finite temperatures since this 
piece of information is crucial for guiding the experimental searches of technologically relevant materials. 
A total of $17$ alloys are predicted to be vibrationally stable, half-metallic, and magnetically ordered at 
room temperature, with total magnetic moments of $2$ and $4$~$\mu_{\rm B}$ and semiconductor band gaps in the 
range of $1$--$4$~eV. On the other hand, all the HH alloys that have been identified as anti-ferromagnetic,  
$21$ in total, turn out to be metallic. We have also found $2$ new magnetic semiconductors that exhibit 
high thermodynamic stability and Curie temperatures. After analyzing the mixing stability of the vibrationally
well-behaved HH alloys, we have identified the following compounds as overall most promising for spintronics 
applications: LiVSi (SC), LiVGe (SC), NaVSi (HM), RbVTe (HM), CsVS (HM), CsVSe (HM) and RbNbTe (HM). On 
the other hand, we have argued that simple doping strategies may be used to improve the mixing stability 
of some of the discarded HH half-metals. Hence, we hope that our computational study will stimulate new 
experimental efforts leading to progress in the field of spin-based electronics.

\section*{Acknowledgments}
This research was supported under the Australian Research Council's Future Fellowship funding scheme (No. FT140100135). 
M. A. S. acknowledges financial support from the Higher Education Commission (HEC) of Pakistan under the IRSIP scholarship 
(PIN:IRSIP 35 PSc 11). Computational resources and technical assistance were provided by the Australian Government and 
the Government of Western Australia through the National Computational Infrastructure (NCI) and Magnus under the National 
Computational Merit Allocation Scheme and The Pawsey Supercomputing Centre.

\end{document}